\documentstyle[aps,preprint,eqsecnum,prd,epsfig]{revtex}

\newcommand{\beq}{\begin{equation}}
\newcommand{\eeq}{\end{equation}}
\newcommand{\bra}[1]{{\langle {#1} |}}
\newcommand{\ket}[1]{{| {#1} \rangle}} 
\newcommand{\vev}{{\langle\,\, \phi\,\, \rangle}}
\newcommand{\vevphi}{{\langle\ \phi\ (\varphi)\ \rangle}}
\newcommand{\vevrphi}{{\langle\ \phi\ (r, \varphi)\ \rangle}}

\newcommand{\adjoint}[1]{{{#1}^\dagger}}

\def\Lag{ {\cal L} }

\def\Id{ {\mbox{1\hskip-0.22em\relax l}}\,} 
\newcommand{\sfrac}[2]{{\textstyle\frac{#1}{#2}}} 
\newcommand{\commut}[1]{ \hbox{\bf [} #1 \hbox{\bf ]} } 

\newcommand{\remove}[1]{}

\newcommand {\comment}[1]{}             

\begin{document}
{\tighten
\preprint{\vbox{\hbox{EUPC/00--01}}}

\title{Charge Violation and Alice Behavior in Global and Textured Strings}

\author{Katherine M.~Benson}
\address{Department of Physics\\
Emory University\\  1510 Clifton Road, NE\\Atlanta, GA\ 30322-2430}

\bigskip
\date{\today}

\maketitle
\begin{abstract}
Spontaneous breaking of global symmetries can produce ``Alice''
strings: line defects which make unbroken symmetries multivalued,
induce apparent charge violation via Aharonov-Bohm interactions, and
form point defects when twisted into loops. We demonstrate this
behavior for both divergent and textured global Alice strings.  Both
adiabatically scatter charged particles via effective Wilson lines.
For textured Alice strings, such Wilson lines occur at all
radii, and are multivalued only inside the string. This produces 
measurable effects,  including path-dependent
 charge violation.
\end{abstract}

\pacs{13.20.He, 12.38.Bx, 13.20.Fc, 13.30.Ce}
}
\narrowtext

\section{Introduction}

Among the defects created in symmetry-breaking phase transitions are
Alice strings.\cite{oldAlice,stringzm,newAlice} Like monopoles
\cite{moncolor}, Alice strings obstruct the global extension of 
unbroken symmetries, making them multivalued when parallel
transported around the string. This algebraic obstruction has two
prominent physical consequences. First, it produces nonconservation of
the associated charges under Aharonov-Bohm scattering around the
string. Second, it induces topologically stable point defects,
which are themselves twisted loops of Alice string.

These Alice features were first described for gauged strings, where a
charged condensate winds asymptotically due to gauge flux on the
string. The gauge flux acts asymptotically on all charged particles in
the same way, through the Wilson line $U(\varphi)$. Thus the
condensate's winding, which makes Alice symmetries multivalued, is
communicated to all charged particles circumnavigating the
string. This communication induces both results above: Aharonov-Bohm
scattering around the string changes both charge and monopole
number. Loops of string, which leave charge and monopole number
well-defined asymptotically, thus support deposited charge (``Cheshire
charge'') and deposited monopole number. \cite{oldAlice,stringzm} We
have recently established, through topological arguments, that Alice
loops can always carry such deposited monopole number by
twisting. \cite{bigauge}

More recent work confirms that Alice phenomena also occur for global
strings. \cite{skyrme,mcgraw,framedrag} Such strings arise when global
symmetries --- describing flavor, chiral, or condensed matter
degeneracies --- break down. Two types of global strings can form:
divergent strings, which wind nontrivially far from the string and
thus have logarithmically divergent tension; and textured strings,
which approach constant field values asymptotically, but carry
topological winding in their finite tension cores.

We note here that divergent global strings inherit their Alice
characteristics from the gauged and textured cases. Their topological
structure coincides with the gauged case: therefore the same
topological arguments identify twisted Alice loops with global
monopoles. \cite{bigauge} Their couplings to fundamental charged
particles are the same as those of their textured cousins. Our careful
Aharonov-Bohm analysis of these interactions shows that divergent
strings induce an effective Wilson line for point charges, usually
equivalent to the Wilson line for the gauged case. Thus divergent
global Alice strings possess the entire constellation of Alice
features: charge-altering Aharonov-Bohm scattering (first noted in
\cite{framedrag}, \cite{mcgraw}), as well as twisted loops which are
point defects.

Recently Benson, Manohar and Saadi found
textured global strings with the same features. \cite{skyrme} At
finite radius, their strings alter charges of particles that
Aharonov-Bohm scatter around them. They also form twisted
loops which are skyrmions, just as gauged Alice strings form twisted
loops which are monopoles.

We show here that these Alice properties emerge generically for textured
Alice strings. The result is striking, as it stems from very different
physics than that of the the gauged case.  The topology differs:
gauged strings have nontrivial first homotopy group $\pi_1$, and form
twisted loops with nontrivial $\pi_2$; while textured strings have
nontrivial $\pi_2$, and form twisted loops with nontrivial $\pi_3$.
Gauged strings vary nontrivially at infinite radius, yielding
long-range flux-mediated interactions; while textured strings are
asymptotically constant, yielding short-range direct interactions.
Yet both are line defects with the Alice constellation of features:
they induce apparent charge violation in Aharonov-Bohm scattering of
charged particles (at least at finite scattering distances), and they
form twisted loops that are topological point defects.

We demonstrate these results below as follows. The divergent global
Alice string has the simplest structure, and thus appears in section
\ref{div}.  Here we construct the divergent Alice string, assuming
that the string couples naturally to fundamental charges via effective
mass terms.  We then show how global Aharonov-Bohm scattering converts
such couplings into an effective charge-violating Wilson line related
to that of the gauged case. This analysis generalizes earlier work of
McGraw \cite{mcgraw}, and lets us  develop tools for studying
textured string scattering in a simpler context.

We then consider the textured Alice string. We construct it
topologically in section \ref{build}, again assuming representations
with effective mass couplings to the fundamental charges. Section
\ref{scatter} explores the scattering of point charges in the textured
string background. This scattering --- which generically mixes
degenerate states through a nonabelian Aharonov-Bohm effect --- has a
rich structure. It induces an effective Wilson line $U(r,\varphi)$
which is radially dependent: while trivial at infinite radius, it
becomes a multivalued group rotation at finite radii. This induces
multiple effects.  At finite radius, eigenstates of locally unbroken
charge alter their charge, in an $r$-dependent way, in
circumnavigating the string.  Even asymptotic unbroken charges can
interact non-trivially with the string, by approaching it,
circumnavigating it at finite radius $r_o$, and then returning with altered charge. 

Finally, in section \ref{twist}, we discuss the connection ---
inherited from the Hopf fibration --- between twisted textured strings
and skyrmions (or textures). We note that the manner in which the
string twists in internal space along the string loop is different
from the gauged case: while the gauged case involves a string whose
plane of rotation rotates along the loop, the textured string rotates
always in the same plane, but with an offset angle increasing along
the loop.

We review our conclusions in section \ref{conclude}.
Intuition-building examples are developed throughout the text,
including the original Alice symmetry-breaking pattern proposed by
Schwarz \cite{oldAlice}, patterns realizable when a larger $SO(6)$
symmetry breaks down, and the Benson-Manohar-Saadi textured string
involving Majorana interactions with point charges. Of these, the
first is most testable, as the global symmetry-breaking transition
responsible for the nematic phase in liquid crystals; the others are
motivated by particle physics GUT phase transitions.

\section{Divergent Global Alice Strings}\label{div}

Consider a global symmetry group $G$ broken to a residual subgroup $H$
by some condensate $\vev$. Divergent strings form under the same
topological conditions as gauged strings: whenever the vacuum manifold
$G/H$ of degenerate $\vev$ orientations has nontrivial loops $\pi_1
(G/H)$. Such loops determine the string's topologically nontrivial
boundary conditions: that is, the winding of $\vev$ asymptotically. We
parametrize this winding via the angle-dependent $G$ rotation
$U(\varphi)$ which generates it:

\beq\vevphi = U(\varphi)\ \vev_o\ . \eeq

Here $U(\varphi)$ acts on $\vev_o$ according to its group
representation.  Recall that, in the gauged case, $U(\varphi)$ is the
Wilson line due to the string's nonabelian flux. Here, it is the
same group rotation, inducing the same condensate winding, but with 
divergent gradient energy cost per unit length.

By the exact sequence for $\pi_1 (G/H)$, $U(\varphi)$ has two
topologically nontrivial forms: it can be a nontrivial loop in $G$,
which cannot lie wholly in $H$; or it can be a path in $G$, ending
on some element $h_o$ in a disconnected component of $H$.

To find an Alice string, which makes some $H$ generators multivalued,
we must adopt the second possibility. This is because the symmetries
$H$ leaving $\vevphi$ invariant, and their generators $T_h$, are
parallel transported around the string:
\begin{eqnarray}
H\ (\varphi) &=& U\ (\varphi) \ \ H_o \ \ U^{-1}\ (\varphi) \nonumber\\
T_h\ (\varphi) &=& U\ (\varphi) \ \ T_{h,o} \ \ U^{-1}\ (\varphi) .
\label{parallelt}
\end{eqnarray}
Thus $T_h \ (\varphi)$ becomes multivalued only when a nontrivial
$U(2\pi)$ fails to commute with $T_{h,o}$.

Thus divergent strings are Alice, with multivalued generators $T_{h}\
(\varphi)$ , whenever $U(2\pi)$ is an element $h_o$ which 1) lies in a
disconnected component of $H_o$, and 2) doesn't commute with Alice
generators $T_{h,o}$. We note further that $U\ (\varphi)$ rotates the
condensate, and connects disconnected components of $H$, so that it
must be generated, at least partially, by broken symmetry generators
$T_b$. To minimize string energy, we take $U\ (\varphi)$ to be
generated uniformly by a single broken generator $T_b$:
\begin{eqnarray}
U\ (\varphi) &=& e^{-i\varphi\, T_b}\nonumber\\
&=& h_0 \ \ \mbox{at}\ \varphi = 2\pi\ .
\label{divuphi}
\end{eqnarray}

Given this construction of the divergent global Alice string,
existence of global monopoles formed by twisted string loops is
guaranteed by \cite{bigauge}. Charge violation due to Aharonov-Bohm
scattering requires more careful analysis. Fundamental fermions $\psi$
carry charges 
\beq Q_h = \int\ d^3x\ \adjoint{\psi}\, T_h\, \psi \eeq
and interact with the condensate $\phi$ through effective mass terms
of either Dirac or Majorana form:
\beq
\Lag_{int} = \left\{\ \begin{array}{ll} 
-m\bar{\psi}\ \phi\ \psi + {\it h.c.} \hspace{1.2in}&\mbox{Dirac}\\
-m\bar{\psi}\ \phi\ \psi_c + {\it h.c.} &\mbox{Majorana}\end{array}
\right.\ \ .
\label{phipsiints}
\eeq
Here the condensate field $\phi$ is in a group representation
transforming either as
\beq
\phi \rightarrow   \left\{\ \begin{array}{ll} 
g\ \phi\ \adjoint{g} \hspace{1.2in}&\mbox{Dirac}\\
g\ \phi\ g^T &\mbox{Majorana}\end{array}\right.\ \ .
\label{phitrans}
\eeq
The Dirac mass term is simpler and generalizes readily to charged
bosons, so we focus on it here. The Majorana interaction arises in
some symmetry breakdowns of $SU(n)$ groups, as for the textured global
string model of Benson, Manohar and Saadi. We return to its role in
Aharonov-Bohm scattering in section \ref{scatter}.

For the Dirac-type interaction, the fermion $\psi$ has a spatially
varying effective mass matrix 
\beq {\cal M} = U\ (\varphi) \ \ \vev_o\ \ U^{-1}\ (\varphi) \eeq 
outside the global string. This determines
the fermion's behavior in Aharonov-Bohm scattering around the
string. For, under Aharonov-Bohm scattering, a fermion in mass
eigenstate $\psi_a$ does three things: 1) it tracks the local mass
eigenstate $\psi_a (\varphi)$; 2) it acquires the time evolution phase
$e^{-im_at}$; and 3) it acquires a ``Berry's phase.'' This Berry's
phase is generally nonabelian, as the residual symmetries $H$ induce
fermion mass degeneracies.  Thus $\psi_a\ (\varphi) $ mixes with other
degenerate eigenstates $\psi_b\ (\varphi)$, according to the nonabelian Wilson
line\cite{wilczekzee}
\beq
W(\varphi) = e^{-\, \int^\varphi\ d\varphi'\ A_{\varphi'}}
\eeq
with the (matrix) gauge field $A_{\varphi'}$ given by
\beq
(A_\varphi)_{ab} = 
\langle\ \psi_b\ (\varphi)\ | \ \partial_\varphi  \ |\ \psi_a\ 
(\varphi)\ \rangle  \ \ 
\label{nonaBphase}
\eeq
in the degenerate subspace.
Altogether, then, a fermion $\psi_a (\varphi = 0)$ becomes,
after circumnavigating the string,
\beq \psi_a' = e^{-im_at}\ W(2\pi)\ \psi_a\,(2\pi)\ \ .\eeq

For the divergent Alice string, the spatially varying mass
eigenstates are
\beq\psi_a\ (\varphi) = U\ (\varphi)\ \psi_a\ \ ,
\label{divmeigens}\eeq
where $\psi_a$ is an eigenstate of the matrix $\vev_o$.  Since $U\
(2\pi) = h_o$, the mass eigenstate $\psi_a\ (\varphi)$ is multivalued
whenever $h_o$ acts nontrivially on $\psi_a$. (Typically, such
multivaluedness consists only of multiplying $\psi_a$ by a 
phase --- for a $Z_2$ Alice string,  a minus sign.)

We now consider the nonabelian Berry's phase for our minimal energy
divergent Alice string (\ref{divuphi}). For divergent
strings, this Berry's phase is either trivial, or cancels any
non-intrinsic multivaluedness in our definition of the mass eigenstate
$\psi_a\ (\varphi)$. (For example, setting $\psi_a\ (\varphi) =
e^{-i\varphi/2}\ \psi_a$, an unphysical multivaluedness, induces
compensating Berry's phase $W(\varphi) = e^{+i\varphi/2}$, showing the
physical mass eigenstates to be truly constant.) Because the
nonabelian Berry's phase will have much richer and more subtle
behavior for textured Alice strings,  we introduce it
carefully here.

For our minimal divergent Alice string (\ref{divuphi}), equation 
(\ref{nonaBphase}) gives a Berry
phase determined by
\beq
(A_\varphi)_{ab} = -i\ 
\langle\ \psi_b\ (\varphi)\ | \ T_b\ |\ \psi_a\ (\varphi)
\ \rangle  =  -i\ 
\langle\ \psi_b\ | \ T_b\ |\ \psi_a
\ \rangle \ \ 
\eeq
using (\ref{divmeigens}). Note that $T_b$ is the broken symmetry
generator generating $U(\varphi)$ --- and thus both string and mass
eigenstate winding --- while the matrix element is taken between
degenerate states. 

Mathematically, the broken symmetry generator $T_b$ can act on state
$|\ \psi_a \ \rangle $ in one of three ways. First it can annihilate
it, leaving the mass eigenstate (\ref{divmeigens}) spatially invariant
with no Berry's phase above. Second, and most generically, it can
rotate $|\ \psi_a \ \rangle $ out of the degenerate mass
subspace. (Note that unbroken generators rotate $|\ \psi_a \ \rangle $
into degenerate states, while broken generators typically rotate $|\
\psi_a \ \rangle $ into nondegenerate ones.) In this case, $ T_b\ |\
\psi_a \ \rangle $ has no component within the degenerate subspace, so
the Berry's phase above vanishes. Here the mass eigenstate $\psi_a\
(\varphi)$ does vary spatially, and any apparent multivaluedness is
physical. In the third case, the rotated state $T_b\ |\ \psi_a \
\rangle $ has components in the degenerate subspace. By basis choice
we diagonalize $T_b$ within the degenerate subspace, taking $\ T_b\ |\
\psi_a \ \rangle  \propto |\ \psi_a \ \rangle $ there. This produces
an overall numeric phase in Berry's phase $W(\varphi)$ , one that exactly
cancels an overall numeric phase in $\psi_a\ (\varphi) = U(\varphi)\
\psi_a$. This shows that $T_b$'s action in the degenerate subspace is
spurious, causing a nonphysical phase variation in $\psi_a\ (\varphi)$
which is eliminated via the self-consistent Berry's phase approach.

Thus, in circumnavigating the divergent global string, fermion mass
eigenstates are always acted on by the effective Wilson loop $W(2\pi)\
U(2\pi)$. Here $U(2\pi)$ is the Wilson loop for the gauge
case. $W(2\pi)$, the nonabelian Berry's phase, is nontrivial {\em only} if
the (broken) string generator $T_b$ fails to rotate the mass
eigenstate entirely out of its degenerate subspace. In that uncommon
case, $W(2\pi)$ cancels any nonphysical phase which creeps into the Wilson
loop $U(2\pi)$.

Behind our calculation of fermion scattering lies the physical
question, ``Is charge conservation violated in Aharonov-Bohm
scattering around the divergent global Alice string?'' Details are
model-dependent, but the mass eigenstate scattering above lays the
foundation for possible charge violation. This is because the
effective Wilson loop, $W(2\pi)\ U(2\pi) = W(2\pi)\ h_o$, makes only
{\em some} mass eigenstates multivalued, while others remain
single-valued. Thus a charge eigenstate
\beq
\ket{q} = \sum_{\rm single}\ c_a\ \ket{\psi_a} +  \sum_{\rm multi}\ c_b\ \ket{\psi_b}\ \ ,
\eeq
where the first sum ranges over single-valued mass eigenstates, the
second over multivalued ones, becomes 
\beq
\ket{q}' = \sum_{\rm single}\ c_a\ \ket{\psi_a} +  \sum_{\rm multi}\ c_b'\ \ket{\psi_b}\ \ ,
\eeq
after circumnavigating the string. Typically, $c_b'$ differ from $c_b$ by a phase. Thus, so long as the charge eigenstate
has both single- and multivalued mass eigenstate components, the
scattered state is physically distinct, and carries distinct charge,
from the initial $\ket{q}$. This means that charges alter, just as in the gauged
case, when Aharonov Bohm scattered around the divergent global Alice
string. In the usual case where $W(2\pi) = \Id$, it is precisely those
charges associated with Alice generators $T_{ho}$ --- which do not
commute with the Wilson loop $U(2\pi)$ --- which may be altered. For in
this case, the effective Wilson loop is just the gauge Wilson loop, $U(2\pi)$, and the Alice charge eigenstates obey
\begin{eqnarray}
T_{ho}\ \ket{q} &=& q\ \ket{q}\nonumber\\
T_{ho}\ \ket{q}' = T_{ho}\ U(2\pi)\ \ket{q} &= & q\ \ket{q}' +\commut{ T_{ho},  U(2\pi)} \ \ket{q}'\ \ ,
\label{wilsonq}
\end{eqnarray}
Thus, in this usual
case, all Alice charges not annihilated by $\commut{ T_{ho},  U(2\pi)}$ 
are nonconserved in scattering around the
divergent global Alice string.

\subsection{The Canonical Divergent Global Alice String}\label{canon}

We illustrate this general formalism with the simplest example, the
Schwarz Alice string, whose scattering in the divergent global string
case was considered by McGraw. \cite{mcgraw} Here $G$ is $SO(3)$, and the
Higgs $\phi$ transforms as a matrix and interacts with fundamental
fermions through Dirac mass terms. $\phi$ develops the vev
\beq\vev = {\rm diag}\ (1,1,-2)\ \ ,
\eeq
with an $O(2)$ residual symmetry $H$ containing  z-rotations $R_z\,(\alpha)$ and the discrete symmetry element
\beq h_o = R_x\, (\pi) = {\rm diag}\ (1, -1, -1)\ \ .\eeq
$\pi_1(G/H) = Z_2$ so we have global $Z_2$ strings generated by 
\beq U(\varphi) = R_x\, (\varphi/2)\eeq
with $U(2\pi) = h_o$. This string is Alice, as $U(2\pi)$ fails to commute
with the unbroken symmetry generator $T_z$; in fact, on parallel
transport around the string,
\beq T_z \rightarrow U\ (2\pi) \ \ T_z \ \ U^{-1}\ (2\pi) = -T_z \ \ .\eeq

We now show that this formal Alice property yields physical Alice
behavior --- with twisted Alice string loops forming monopoles, and
charge-violating Aharonov Bohm scattering. As always for divergent
global strings, twisted loops form monopoles due to the arguments of
\cite{bigauge}. Charge violation occurs because some fermion mass eigenstates
are double-valued; in particular, we have spatially varying mass
eigenstates
\begin{eqnarray}
\{\ \ket{\psi_a\ (\varphi)}\ \} &=& R_x\, (\varphi/2)\ \ \left\{\ \left(\begin{array}{c}
1\\[-8pt] \\[-8pt] \  \end{array} \right), \left(\begin{array}{c}
\\[-8pt] 1\\[-8pt] \ \end{array} \right),  \left(\begin{array}{c}
\\[-8pt] \\[-8pt] 1\end{array} \right)\ \right\} \nonumber\\[10pt]
&=& \left\{\ \left(\begin{array}{c}
1\\[-8pt] \phantom{ \cos\ (\varphi/2)} \\[-8pt] \phantom{ \cos\ (\varphi/2)} \end{array} \right), \left(\begin{array}{c}
\\[-8pt] \cos\ (\varphi/2)\\[-8pt]  \sin\ (\varphi/2)\end{array} \right),  \left(\begin{array}{r}
\\[-8pt] -\sin\ (\varphi/2) \\[-8pt]  \cos\ (\varphi/2)\end{array} \right)\ \right\}\ \ .
\end{eqnarray}
Note that the first two mass eigenstates are degenerate, while the
last two are double-valued. The nonabelian Berry's phase vanishes, as
the string generator $T_x$ does not connect any degenerate mass eigenstates. Thus this example has the same effective Wilson loop $U(2\pi)$ as in the gauged case, making two of three mass eigenstates double-valued. As expected, fermions reverse charge in circumnavigating the global string, for charge eigenstates
\beq
\ket { \pm} = \frac{1}{\sqrt{2}}\ \left(\begin{array}{r}1\\[-8pt] \pm i\\[-8pt] \ \end{array}\right) \hspace{1.2in} 
\ket{0} = \ \left(\begin{array}{r}\\[-8pt] \\[-8pt] 1 \end{array}\right) 
\eeq
scatter as $\ket{\pm} \rightarrow \ket{\mp}, \ket{0} \rightarrow -
\ket{0}$ in traversing the string. This can be seen directly from the
action of the effective Wilson loop $U(2\pi) = h_o$ on the charge
eigenstates, or from the underlying spatial variation of the
constituent mass eigenstates.

\section{Building Textured Alice Strings}\label{build}

Again consider the breaking of a global symmetry $G$ to residual
subgroup $H$ by Higgs condensate $\vev$. Finite tension strings form
only if they are textured; that is, if they approach a constant vev
$\vev_o$ asymptotically and carry topological winding in their
cores. Such winding occurs whenever the vacuum manifold has nontrivial
homotopy $\pi_2(G/H)$. This describes the condensate's nontrivial
variation through degenerate vacuum states in the plane transverse to
the string; because each point at spatial infinity approaches the same
vacuum state $\vev_o$, the transverse plane becomes identified with a
sphere $S^2$, and has topologically nontrivial vacuum configurations
classified by $\pi_2(G/H)$.

We display the condensate's spatial variation in Figure 1, noting that
it is generated by some group rotation $U(r, \varphi)$ acting on the
single asymptotic vev $\vev_o$: 
\beq \vevrphi = U(r, \varphi)\ \vev_o\ . \eeq 
$U(r,\varphi)$ acts on $\vev_o$ according to its group
representation, and varies over $G$ in some single-valued, nonsingular
way. By the exact sequence for $\pi_2 (G/H)$, $U(r, \varphi)$ has two
topologically nontrivial forms: it can be a nontrivial map in $G$,
which cannot lie wholly in $H$; or it can be a map in $G$, ending
asymptotically on a nontrivial loop $h(\varphi)$ in $H$. Again, to
find Alice behavior, we restrict ourselves to the second case, where
$U(r, \varphi)$ approaches a nontrivial loop $h(\varphi)$ in H at
large $r$. (Thus $H$ must have nontrivial $\pi_1$.) Remarkably, even
though the map $U(r,\varphi)$ is single-valued, it can cause charges
to become multivalued, or Alice, as we shall see by construction.

\vspace{8pt}
\epsfig{file=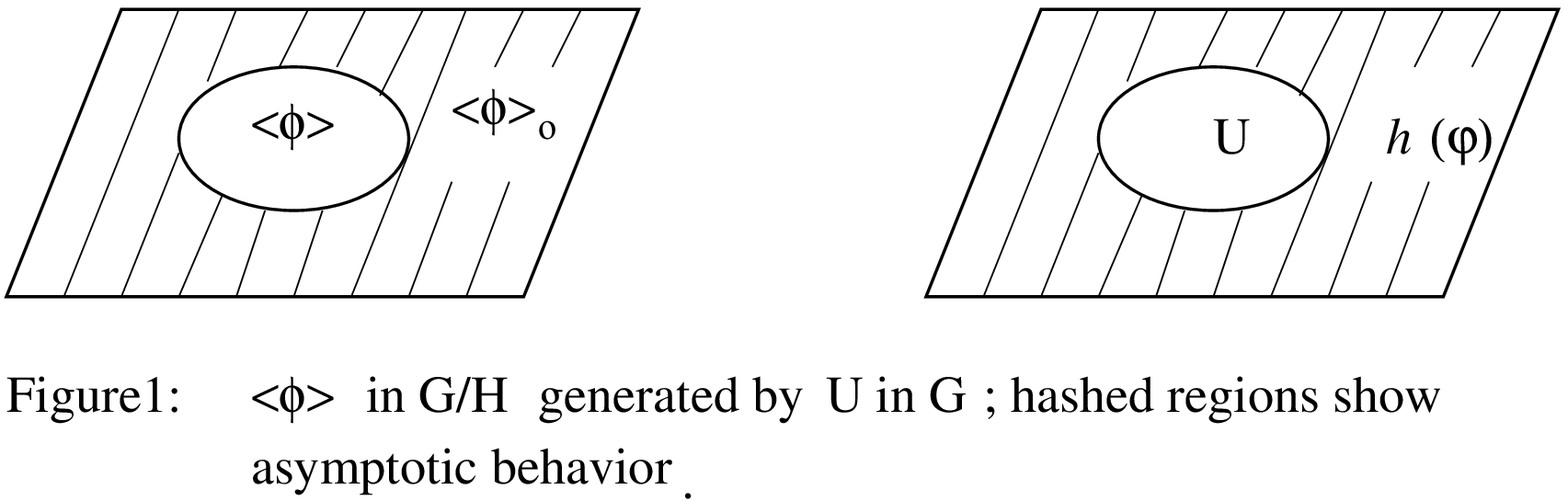, width=5.5in}

Construction could, of course, vary greatly by model; here, however,
we suggest a useful ansatz. Say our nontrivial loop $h(\varphi)$ is
generated by unbroken generator $T_h$, which anticommutes with some
group element $b_o$. (This happens whenever $G$ contains $SU(2)$ or $SO(3)$ subgroups, but could occur for more exotic
symmetry breaking patterns as well.) Taking $b_o$ to be generated by
broken generator $T_b$, 
\beq b_o = e^{-iF_o T_b} \hspace{1.2in}h(\varphi) = e^{-i\varphi T_h}\ \ ,\eeq 
a nontrivial map $U(r,\varphi)$ is given by
\beq
U(r,\varphi) = h(\varphi/2)\ \ e^{-iF(r)\, T_b}\ \ h(\varphi/2)\hspace{0.75in}
\mbox{where} \quad F(r) \rightarrow \left\{ \begin{array}{ll}0&r\rightarrow\infty\\F_o&r\rightarrow 0
\end{array}\right.
\label{genU}
\eeq
Note that $U = b_o$ at the origin, since $b_o h = h^{-1} b_o$, and
approaches the nontrivial loop $h(\varphi)$ at infinite radius. The
only caveat is that $U(r,\varphi)$ must be single-valued. This could
occur if $h(\pi)$ commutes with $T_b$, a possibility we exploit in the canonical Alice model. For our GUT-inspired models, however, we insure
single-valuedness by taking $T_b$ and $T_h$ to lay within a partially
broken $SU(2)$ subalgebra. We then choose Lie algebra bases so that the generators $T_b$ and $T_h$ anticommute, with 
\beq e^{-iF(r)\, T_b} = \cos\,F(r)\ \Id - i\sin\, F(r) \ T_b \eeq
within the subspace where $T_b$ and $T_h$ are nontrivial. (We note
that we can make these choices whenever $G$ includes $SU(N)$ factors,
for $N\ge 2$, or $SO(N)$ factors, for $N\ge 4$, as we show by
construction below). Thus, for these models,
\beq
U(r,\varphi) = h(\varphi/2)\ \ e^{-iF(r)\, T_b}\ \ h(\varphi/2) \quad = \quad
\cos \, F(r) \ h(\varphi) - i\sin\, F(r)\ T_b\ \ ,
\eeq
within the subspace where $T_b$ and $T_h$ are nontrivial.  This gives
the loop $h(\varphi)$ at $r\rightarrow\infty$, and anticommuting
broken group element $b_o$ at the origin, where $F =F_o =
\pi/2$. Moreover, this map is single-valued and nonsingular
everywhere, and so acts on $\vev_o$ to create a nontrivial textured
string whenever the loop $h(\varphi)$ is nontrivial in $H$.

Thus far, our ansatz constructs a nontrivial textured string. When is
this string Alice; that is, when does an unbroken generator $T_{h'}\
(r)$ determine non-conserved charge eigenstates in circumnavigating
the string?  Surprisingly, the answer is nontrivial,
and depends on the nonabelian Berry phase to resolve
ambiguities. Formally, unbroken generators are parallel transported
throughout the textured string:
\beq
T_h\ (r, \varphi) = U\ (r, \varphi) \ \ T_{h,o} \ \ U^{-1}\ (r,\varphi) .
\eeq
Naively, since $U(r,2\pi) =U(r,0)$, all generators are single-valued;
none would appear to be Alice. Closer analysis, however, in all our
examples below and in section \ref{scatter}, reveals that {\bf any}
generator $T_{h'}$ which fails to commute with $ U\ (r, \varphi)$, at {\bf any} angle $\varphi$, is Alice: its
charges alter when Aharonov-Bohm scattered around the string at finite,
nonzero radius.

We construct textured Alice strings in three models below, following
this ansatz. In the first, a symmetry breakdown of $SO(N)$, the
condensate $\vev$ interacts with charged fermions through a Dirac mass
coupling. The second, the Benson, Manohar, Saadi textured string,
involves $SU(N)$ symmetry breakdown and Majorana mass
couplings. Finally, we construct a textured Alice string in the
canonical Schwarz model itself, which exploits subtleties of the
embedding of $O(2)$ in $SO(3)$.

\subsection{$SO(N)$ symmetry breakdown and the $SO(6)$ model}\label{SON}

Let the unbroken symmetry group $G$ be $SO(N)$, with Higgs $\phi$
transforming in the adjoint and coupling to fermions $\psi$ through
Dirac mass terms (equations (\ref{phitrans}) and (\ref{phipsiints})).
For $N\ge 4$, we can construct a textured string varying over an
$SO(4)$ subspace, which makes some unbroken generators
$T_{h'}$ Alice; we show the case with $N=6$ for definiteness.

$\phi$ develops the vev 
\beq \vev = {\rm diag}(1,1,1,-1,-1,-1)\eeq
with residual
symmetry $H = SO(3)\ \times\ SO(3)\ \times\ Z_2$ of rotations in the
upper left $3\times 3$ block, rotations in the lower right $3\times 3$
block, and the discrete element $-\Id$. Now consider the $4\times 4$
subspace of rows and columns 2 through 5, where $\phi$ assumes values
\beq \vev_\Box = {\rm diag}(1,1,-1,-1)\ \ .\eeq 
It is in this subspace, of an $SO(4)$ theory
broken to $SO(2)\ \times \ SO(2)$, where our textured string will vary. We
therefore reindex these rows from 1 to 4 in what follows.

A natural basis for $so(4)$ in our subspace is the set $\{\ T_{ij}\ \}$
generating rotations in the six planes $ij$. 
However we exploit the basis 
\beq
\begin{array}{rrr}
J_1 = T_{12} + T_{34} & \quad J_2 =  T_{23} + T_{14} &\quad J_3 = T_{13} - T_{24}\\
K_1 = T_{12} - T_{34} & K_2 =  T_{23} - T_{14} & K_3 = T_{13} + T_{24}
\end{array} \ \ ,
\eeq
which makes the isomorphism $SO(4) = SU(2) \ \times\ SU(2)$ clear:
\beq
\commut{J_i, J_j} = 2i\, \epsilon_{ijk}\ J_k \hspace{0.6in}
\commut{K_i, K_j} = 2i\, \epsilon_{ijk}\ K_k \hspace{0.6in}
\commut{J_i, K_j} = 0\ \ .
\label{su2su2}
\eeq
Moreover, the sets $\{J_i\}$ and $\{K_i\}$ truly behave like Pauli
matrices: anticommuting among themselves, and squaring to $\Id$ within
the 4-dimensional subspace, so that
\beq e^{-i\alpha\, J_i} = \cos\,\alpha\ \Id - i\sin\, \alpha \ J_i 
\label{pauliexp}
\eeq
within the $\Box$-subspace.

Now consider the $so(3)$ subalgebra of $g$ generated by $\{\, J_i\, \}$.
Only $J_1$ remains unbroken by the vev $\vev_\Box$; it generates the nontrivial loop
\beq h(\varphi) = e^{-i\varphi\ J_1} \eeq
in our full unbroken symmetry group $SO(3)\ \times\ SO(3)\ \times Z_2$. Since $J_2$ anticommutes with $J_1$, the map 
\beq
U\ (r,\varphi) = h(\varphi/2)\ \ e^{-iF(r)\, J_2}\ \ h(\varphi/2) \quad = \quad
\cos \, F(r) \ h(\varphi) - i\sin\, F(r)\ J_2\ \ 
\eeq
in the $SO(4)$ subspace, with 
\beq
\quad F(r) \rightarrow \left\{ \begin{array}{ll}0&r\rightarrow\infty\\\pi/2&r\rightarrow 0
\end{array}\right.
\eeq
acts on $\vev_\Box$ to create a nontrivial textured string. 
This is because $U(r,\varphi)$  is single-valued and nonsingular, giving 
$b_o = -iJ_2$ at the origin and 
the nontrivial loop $h(\varphi)$ asymptotically.

Because $\phi$ transforms in the adjoint, the textured string looks like
\beq \vev\ (r,\varphi)  = 
U\ (r,\varphi) \ \vev_o\  U^{-1}\ (r,\varphi) \ \ .
\label{form1}\eeq
We can also use the unbroken symmetry to rewrite this as
\beq
 \vev\ (r,\varphi)  = h(\varphi/2)\ \ e^{-iF(r)\, J_2}\ \ \vev_o\ \ e^{+iF(r)\, J_2}\ \ h^{-1}\ (\varphi/2)
\ \ .\label{form2}\eeq

We see here how ambiguities arise in establishing the string as
Alice. For the first form (\ref{form1}) for $\vev_\Box$ encourages us
to defines spatially varying mass eigenstates as
\beq \psi_a\ (r,\varphi) = U\ (r,\varphi) \ \psi_a \ \ ,\eeq
where $\psi_a$ are eigenstates of $\vev_o$. These would be
single-valued, implying that no charges are Alice. However, the second
form indicates spatially varying mass eigenstates of
\beq \psi_a\ (r,\varphi) = h(\varphi/2)\ \ e^{-iF(r)\, J_2} \ \psi_a \ \ ,\eeq
double-valued for all $\psi_a$ in the four-dimensional
$\Box$-subspace (and single-valued elsewhere, where $h(\varphi/2) = \Id$). 
Thus any charge eigenstates with components
straddling the $\Box$ and non-$\Box$ subspaces would appear to be
Alice. Careful nonabelian Aharonov-Bohm scattering analysis resolves this
ambiguity in section \ref{scatter}, with a physically sensible 
 answer. At the origin (where any angle-dependence would be
singular) and at spatial infinity (where the textured string leaves no
physical trace, since the vev $\vev$ is constant), no charges are
Alice. However, at finite intermediate $r$, all charges whose
generators $T_{h'}$ fail to commute with $U\ (r,\varphi)$ become multivalued, or
Alice, in a curiously $r$-dependent way.

\subsection{$SU(N) \rightarrow SO(N)$; the Benson-Manohar-Saadi string}
\label{Maj}

We here review the Benson-Manohar-Saadi string in simplified formalism. 
Let the unbroken symmetry group $G$ be $SU(N)$, with Higgs $\phi$
transforming as
\beq
\phi \rightarrow   
g\ \phi\ g^T 
\label{phitrans2}
\eeq
and coupling to fermions $\psi$ through
the Majorana  mass term  (\ref{phipsiints}).
For $N\ge 2$, there exists a textured string varying over an
$SU(2)$ subspace, which  makes some unbroken generators
$T_{h'}$ Alice; we show the case with $N=3$ for definiteness.

Here $\phi$ develops the vev 
\beq \vev = \Id \eeq
with residual
symmetry $H = SO(3)$ of orthogonal rotations ($h^T = h^{-1}$).
We take the $SU(3)$ generators to be
\beq
\begin{array}{lcl}
\mbox{symmetric:\ }&\quad& \sigma_{x(12)},\   \sigma_{x(23)},\  \sigma_{x(31)},\   \sigma_{z(12)},\  {\textstyle\frac{1}{\sqrt{3}}}\ {\rm diag}\ (1,1,-2)\\
\mbox{antisymmetric:\ }&& \sigma_{y(12)},\   \sigma_{y(23)},\  \sigma_{y(31)}
\ \ ,
\end{array}\eeq
where $\sigma_{x(ij)},\  \sigma_{y(ij)},\  \sigma_{z(ij)}$ denote Pauli
matrices in the $(ij)$ plane. Under symmetry-breaking, all symmetric
generators are broken, all antisymmetric generators unbroken. (This
statement generalizes to all $N\ge 3$, with unbroken subgroup $SO(N)$.)

We now construct the textured string in the (12) $SU(2)$ subgroup.
Here the unbroken generator $ \sigma_{y(12)}$ generates the nontrivial loop
\beq h(\varphi) = e^{-i\varphi\  \sigma_{y(12)}} \eeq
and the broken generator $\sigma_{z(12)}$ both anticommutes with $ \sigma_{y(12)}$ and obeys
\beq e^{-i\alpha\, \sigma_{z(12)}} = \cos\,\alpha\ \Id - i\sin\, \alpha \ \sigma_{z(12)} \eeq
in the (12) plane. Thus our ansatz gives a nontrivial textured string generated by 
\beq
U\ (r,\varphi) = h(\varphi/2)\ \ e^{-iF(r)\, \sigma_{z(12)}}\ \ h(\varphi/2) \quad = \quad
\cos \, F(r) \ h(\varphi) - i\sin\, F(r)\ \sigma_{z(12)}\ \ 
\eeq
in the (12) plane, with 
\beq
\quad F(r) \rightarrow \left\{ \begin{array}{ll}0&r\rightarrow\infty\\\pi/2&r\rightarrow 0
\end{array} \ \ \right.
\eeq
Again $U(r,\varphi)$  is single-valued and nonsingular, giving 
$b_o = -i\sigma_{z(12)}$ at the origin and 
the nontrivial loop $h(\varphi)$ asymptotically.

Applying the transformation law (\ref{phitrans2}),  the textured string looks like
\beq \vev\ (r,\varphi)  \  = \ 
U\ (r,\varphi) \ \vev_o\  U^{T}\ (r,\varphi) \ \ 
 \  = \  h(\varphi/2)\ \ e^{-2iF(r)\,\sigma_{z(12) }}\ \ h^{-1}\ (\varphi/2)
\ \ .\eeq
Again, the existence of double-valued mass eigenstates in the (12) plane,
\beq 
\psi_\pm\ (\varphi) \  =\  h\, (\varphi/2)\ \ \left\{\ \left(\begin{array}{c}
1\\[-8pt] \  \end{array} \right), \left(\begin{array}{c}
\\[-8pt] 1\end{array} \right) \ \right\}
 \  =\  \left\{\   \left(\begin{array}{c}
 \cos\ (\varphi/2)\\[-8pt]  \sin\ (\varphi/2)\end{array} \right),  \left(\begin{array}{r}
 -\sin\ (\varphi/2) \\[-8pt]  \cos\ (\varphi/2)\end{array} \right)\ \right\}
\label{BMSmasseig}\eeq
makes all (13) and (23) unbroken generators appear to be Alice (since
the mass eigenstate $(0,0,1)$ is spatially invariant).
 But such omnipresent Alice behavior again cannot be
physical --- our solution is nonsingular at the origin and constant at
spatial infinity, and thus cannot physically cause Alice scattering at those radii.

Again we resolve the conundrum in section \ref{scatter}, via careful
nonabelian Aharonov-Bohm scattering analysis. At the origin and at
spatial infinity, we show that no charges are Alice. However, at
finite intermediate $r$, all charges whose generators $T_{h'}$ fail to
commute with $U\ (r,\varphi)$ show $r$-dependent Alice behavior.

\subsection{The Canonical Alice model}\label{canonagain}

Surprisingly, the rich laboratory of the original Schwarz model (or
nematic liquid crystal) also supports a textured Alice string given by
our ansatz. Here, from section \ref{canon}, 
$G$ is $SO(3)$, with  the
Higgs $\phi$ transforming as a matrix and giving Dirac masses to the  fundamental fermions. $\phi$ develops the vev
\beq\vev = {\rm diag}\ (1,1,-2)\ \ ,
\eeq
with an $O(2)$ residual symmetry $H$ containing  z-rotations $R_z\,(\alpha)$ and the discrete symmetry element
\beq h_o = R_x\, (\pi) = {\rm diag}\ (1, -1, -1)\eeq
which anticommutes with unbroken generator $T_z$. 

A topological subtlety arises here. From the exact sequence
for $\pi_2(G/H)$, all nontrivial textured strings are generated by
a map $U(r,\varphi)$ ending on a nontrivial loop $h(\varphi)$ in
$H$. Such a map $U(r,\varphi)$ unwinds the loop in $H$ in the larger
group $G$. For the symmetry breaking at hand, we note that only $O(2)$
loops of even winding are unwindable in $SO(3)$. Thus, the lowest
winding textured string must end on a winding 2 loop in $O(2)$,
\beq
h(\varphi) = R_z\, (2\varphi)\ \ .\eeq

Given this, our ansatz gives a lowest winding textured string, with
\beq
U(r,\varphi) = R_z\, (\varphi)\ \ R_x\, (\ F(r)\ )\ \ R_z\,(\varphi)
\hspace{0.75in}
\mbox{where} \quad F(r) \rightarrow \left\{ \begin{array}{ll}0&r\rightarrow\infty\\ \pi&r\rightarrow 0
\end{array}\right.
\eeq
Note that $U$ is manifestly single-valued. It approaches the
nontrivial loop $R_z\, (2\varphi)$ at infinite radius, and the value
$R_x\, (\pi) = h_o$ at the origin, since $h_o R_z\,(\varphi) =
R_z^{-1}\,(\varphi) h_o$.

$U$ determines the textured string
\beq \vev\ (r,\varphi)  \  = \ 
U\ (r,\varphi) \ \vev_o\  U^{-1}\ (r,\varphi) \ \ 
 \  = \   R_z\, (\varphi)\ \ R_x\, (\ F(r)\ )\ \ \vev_o\ \ R_x^{-1}\, (\ F(r)\ )\ \ R_z^{-1}\, (\varphi)
\ \ .\eeq
This determines apparently single-valued mass eigenstates,
\beq 
\{\ \psi_a\ (\varphi)\ \} \quad =\quad R_z\, (\varphi)\ \ R_x\, (\ F(r)\ )\ \ 
\left\{\ \left(\begin{array}{c}
1\\[-8pt] \\[-8pt] \  \end{array} \right), \left(\begin{array}{c}
\\[-8pt] 1\\[-8pt] \ \end{array} \right),  \left(\begin{array}{c}
\\[-8pt] \\[-8pt] 1\end{array} \right)\ \right\}\ \ .
\eeq
However, degeneracy of the first two eigenstates again forces a
careful nonabelian Aharonov-Bohm analysis, which again reveals
nontrivial Alice behavior at finite radius.

\section{Charge Scattering by Textured Alice Strings}\label{scatter}

We have identified three models with textured strings. Each string is
generated by the action on a vev $\vev_o$ of a single-valued group map
\beq
U(r,\varphi) = h(\varphi/2)\ \ e^{-iF(r)\, T_b}\ \ h(\varphi/2)\hspace{0.75in}
\mbox{where} \quad F(r) \rightarrow \left\{ \begin{array}{ll}0&r\rightarrow\infty\\F_o&r\rightarrow 0
\end{array}\right. \ \ ,
\label{genUagain}
\eeq
with $h(\varphi) = e^{-i\varphi T_h}$ a minimal nontrivial loop in $H$ shrinkable in $G$.

We now consider Aharonov-Bohm scattering of fundamental fermions
$\psi$ by such a string. Details of the scattering analysis differ,
for Dirac and Majorana fermions; however, both give the same result:
charge-violating scattering at finite nonzero radii only. We treat
each in turn.

\subsection{For Dirac Fermions}\label{dirac}

When $\psi$ couples to $\phi$ through Dirac couplings
(\ref{phipsiints}), $\phi$ transforms under $U$ by conjugation and
produces the effective mass matrix 
\beq
 \vev\ (r,\varphi)  = h(\varphi/2)\ \ e^{-iF(r)\, T_b}\ \ \vev_o\ \ e^{+iF(r)\, T_b}\ \ h^{-1}\ (\varphi/2)
\ \ .\label{form2again}\eeq
for $\psi$. Choosing $\vev_o$ diagonal (as in our examples) gives
apparent spatially varying mass eigenstates
\beq \psi_a\ (r,\varphi) = h(\varphi/2)\ \ e^{-iF(r)\, T_b} \ \hat{e}_a \ \ ,
\label{masseig}
\eeq
with $\hat{e}_a$ a unit basis vector. These mass eigenstates may or
may not appear multivalued. First, $T_h$ may annihilate $e^{-iF(r)\,
T_b} \ \hat{e}_a$, giving an angle-independent mass eigenstate
$\psi_a$.  Otherwise, there are two possibilities for $\psi_a$.  If
the unit winding loop in $H$ is shrinkable in $G$ (as for the $SO(6)$
example of section \ref{SON}), the remaining mass eigenstates are 
double-valued. However, if the smallest winding loop in $H$
shrinkable in $G$ has even winding (as for the canonical Alice model
example of section \ref{canonagain}), all mass eigenstates are
single-valued.

However, the relevant object, as discussed for divergent global strings in
section \ref{div}, is not the mass eigenstate (\ref{masseig}) but the
mass eigenstate acted on by nonabelian Berry's phase $W(r, \varphi)$:
\beq
W(r, \varphi) = e^{-\, \int^\varphi\ d\varphi'\ A_{\varphi'}}
\eeq
where
\beq
(A_\varphi)_{ab} = 
\langle\ \psi_b\ (r, \varphi)\ | \ \partial_\varphi  \ |\ \psi_a\ 
(r, \varphi)\ \rangle  \ \ 
\label{nonaBphaseagain}
\eeq
within a subspace of degenerate mass eigenstates $\psi_a\ (r,\varphi)$.
For our mass eigenstates (\ref{masseig}), this is determined by
\begin{eqnarray}
(A_\varphi)_{ab} &= & \frac{-i}{2}\ \bra{e_b}\ \  e^{+iF(r)\, T_b}\ \ T_h \ \ 
 e^{-iF(r)\, T_b}\ \ \ket{e_a} \nonumber\\[4pt]
&=&  \bra{e_b}\ \  \left(\ \cos\ [kF(r)]\ (-iT_h/2) \ \ + \ \ \sin\ [kF(r)]\ (-iT_{b'}/2)\ \right)\ \ \ket{e_a}\ \ .
\label{conjid}
\end{eqnarray}
Here $T_{b'}$ is another broken generator ($J_3$ for the $SO(6)$ case,
$T_y$ for the canonical Alice case); $k$ is 2 for the $SO(6)$
case and $1$ for the canonical Alice case. This simple reduction comes
from special algebraic features of the groups chosen: for $SO(6)$,
from the properties discussed in equations (\ref{su2su2}) through
(\ref{pauliexp}); for the canonical case, from traits of the rotation
group $SO(3)$ and its generators. In both cases, the broken generator
$T_{b'}$ does not connect degenerate mass eigenstates, while the unbroken generator connects only degenerate states. Thus the
nonabelian Berry's phase acts on mass eigenstates as
\beq
W(r, \varphi) = e^{+i\varphi\  \cos\ (kF(r))\ T_h/2}\ \ .
\eeq
Thus, in scattering around the string, mass eigenstates $ \psi_a\ (r,\varphi = 0) =  e^{-iF(r)\, T_b} \ \hat{e}_a$ are acted on by an effective Wilson loop
\beq
W(r, 2\pi)\ h(\pi) = h(\,\pi\, [1- \cos\ (kF(r))\, ]\ )\ \ .\eeq
This gives single-valued eigenstates at spatial infinity, where $F =
0$, and at the origin, where $kF = \pi$ for both models. For finite
$r$, however, all mass eigenstates not annihilated by $T_h$ become
multivalued. Thus we have multivalued mass eigenstates (for the
$SO(6)$ model, those induced by $\hat{e}_a$ in the $\Box$-subspace;
for the canonical Alice model, those induced by $\hat{e}_1$ and
$\hat{e}_2$); and we have single-valued mass eigenstates (for the
$SO(6)$ model, those induced by $\hat{e}_a$ outside the
$\Box$-subspace; for the canonical Alice model, the one induced by
$\hat{e}_3$).  As discussed in the last paragraph preceding section
\ref{canon}, this means that charge eigenstates with both multi-
and single-valued components typically undergo charge-violating Aharonov-Bohm
scattering.

More specifically, consider the angle $\alpha(r) \equiv \pi\, [1-
\cos\ (kF(r))\, ]$. $h(\,\alpha(r)\, )$ truly acts as an r-dependent
Wilson loop, for Aharonov-Bohm scattering around the textured Alice
string. To see consequences of this Wilson loop, we consider two
specific Aharonov-Bohm scattering experiments: in the first, we bring
a charge radially inward from spatial infinity to $r$, circumnavigate
the string, then return radially to spatial infinity to compare or
interfere with untransported asymptotic charges. In the second, an
observer within the texture defines charges, as related to the
unbroken symmetries he observes locally, and observes parallel
transport of those charges around the string. Either case gives the
same criterion for apparent charge violation, and in each, this
criterion is met generically for circumnavigation at finite nonzero
$r$.

\subsubsection{Reference Charges at Infinity}

First consider comparison of charges defined at spatial infinity. Here
we define charge eigenstates $\ket{q_{h'}}$ for all unbroken
generators $T_{h'}$. We note first that the nonabelian Aharonov Bohm
gauge field $A_r$ (analogous to equation (\ref{nonaBphaseagain}))
vanishes for our models. Thus radial paths contribute no Aharonov-Bohm
phase, and mass eigenstates simply track the $r$-dependent eigenstate
(\ref{masseig}) during any radial transit. Thus, on the radial transit
inward, the charge eigenstate $\ket{q_{h'}}$ of generator $T_{h'}$
evolves to 
\beq \ket{q_{h'}}_{{\rm at }\ r} = e^{-iF(r)\, T_b} \ \ 
\ket{q_{h'}}\ \ , \label{transpq}\eeq 
taking $\varphi = 0$. After parallel
transport around the string, the charge eigenstate acquires the
effective Wilson loop $h(\,\alpha(r)\, )$: 
\beq 
\ket{q_{h'}}'_{{\rm at}\ r} = h(\,\alpha(r)\, )\ \ e^{-iF(r)\, T_b} \ \ 
\ket{q_{h'}}\ \ .
\label{circumq}
\eeq 
Now this circumnavigated charge must be transported back to
spatial infinity, for comparison with other asymptotic charges. This
gives for the final transported charge 
\beq 
\ket{q_{h'}}'_{{\rm at }\ \infty} =e^{+iF(r)\, T_b} \ \ 
h(\,\alpha(r)\, )\ \ e^{-iF(r)\, T_b} 
\ \ \ket{q_{h'}}\ \ .  \eeq 
Note we can see this by decomposing our
circumnavigated charge (\ref{circumq}) at $r$ into mass eigenstates $e^{-iF(r)\, T_b}
\ \ \ket{e_a}$ at $r$: 
\beq \ket{q_{h'}}'_{{\rm at }\ r} = \sum_a\
\left( \ e^{-iF(r)\, T_b} \ \ \ket{e_a}\ \right)\  \bra{e_a}\ e^{+iF(r)\,
T_b}\ \ h(\,\alpha(r)\, )\ \ e^{-iF(r)\, T_b} \ \ \ket{q_{h'}}\ \ .
\eeq 
Each mass eigenstate $e^{-iF(r)\, T_b} \ \ \ket{e_a}$ then
adiabatically tracks back to its value $\ket{e_a}$ at $r=\infty$,
giving
\begin{eqnarray}
\ket{q_{h'}}'_{{\rm at }\ \infty} &=&  \sum_a\ \ket{e_a}\  \bra{e_a} 
e^{+iF(r)\, T_b}\ \ h(\,\alpha(r)\, )\ \ e^{-iF(r)\, T_b} \ \ \ket{q_{h'}}
\nonumber\\[4pt]
&=& e^{+iF(r)\, T_b}\ \ h(\,\alpha(r)\, )\ \ e^{-iF(r)\, T_b} \ \ 
\ket{q_{h'}}\ \ .
\end{eqnarray}

Thus, asymptotic symmetry generators $T_{h'}$ define charge eigenstates
$\ket{q_{h'}}$ which acquire the effective Wilson loop $e^{+iF(r)\,
T_b}\ \ h(\,\alpha(r)\, )\ \ e^{-iF(r)\, T_b}$ in our thought
experiment: when brought in to radius $r$, parallel transported about
the string, then returned to spatial infinity.

\subsubsection{Reference Charges at $r$}

Now consider charges as defined by an observer within the texture, at
radius $r$ and angle $\varphi=0$.  Because $\vev$ assumes a
nonasymptotic value at finite $r$, its residual symmetry group $H$ is
not the asymptotic one. In particular, for our $\vev$-solutions
(\ref{form2again}), the unbroken symmetry generators leaving $\vev$ invariant at $r$ are
\beq
 T_{h',r} =  e^{-iF(r)\, T_b} \ \  T_{h'}\ \ e^{+iF(r)\, T_b}
\ \ ,
\eeq
again for $\varphi = 0$. Thus the charge eigenstates at $r$ are given by
\beq \ket{q_{h'}}_{{\rm at }\ r} = e^{-iF(r)\, T_b} \ \ 
\ket{q_{h'}}\ \ , \eeq 
related to the asymptotic charges by change of basis.
Note that this is the same equation as equation (\ref{transpq}) for
asymptotic charge eigenstates parallel transported to $r$; however,
here the meaning is different, as these are the charge eigenstates as
{\em defined} at $r$.

After parallel transport around the string, the charge eigenstate
acquires the effective Wilson loop $h(\,\alpha(r)\, )$: 
\beq
\ket{q_{h'}}'_{{\rm at}\ r} = h(\,\alpha(r)\, )\ \ e^{-iF(r)\, T_b} \
\ \ket{q_{h'}}\ \ .  \eeq
Thus, for an observer within the texture, symmetry generators $ e^{-iF(r)\, T_b} \ T_{h'} \ e^{+iF(r)\, T_b}$ define charge eigenstates $e^{-iF(r)\, T_b} \ \ket{q_{h'}}$, which acquire the Wilson loop
$\ h(\,\alpha(r)\, )$ when  parallel
transported about the string.

\vspace{0.25in}

We thus see that our two experiments, involving asymptotic or
$r$-defined charge, are related by change of basis, and give the
same criterion for charge nonconservation.  Recall that any charge
$q_{h'}$ is nonconserved if 1) its associated generator fails to
commute with the Wilson loop; and 2) the charge eigenstate is not
annihilated by the nontrivial commutator, as discussed in equation
(\ref{wilsonq}). Here that means charge $q_{h'}$ will be nonconserved, in either thought experiment, if
\beq
\commut{ \ T_{h'}, \  
e^{+iF(r)\, T_b}\ \ h(\,\alpha(r)\, )\ \ e^{-iF(r)\, T_b}\ 
} \ \ket{q_{h'}}\ \ \ne \ \ 0
\eeq
or equivalently, for finite nonzero $r$,
\beq
\commut{ \ T_{h'}, \  
e^{+iF(r)\, T_b}\ \ T_h\ \ e^{-iF(r)\, T_b}\ 
} \ \ket{q_{h'}}\ \ \ne \ \ 0
\eeq

Recalling the algebraic result (\ref{conjid}) for our models, our criterion for $q_{h'}$ charge violation becomes
\beq
\commut{\ T_{h'}, \ \left(\ \cos\ [kF(r)]\ T_h \ \ + \ \ \sin\ [kF(r)]\ T_{b'}\ \right)\ }  \ \ket{q_{h'}}\ \ne 0\ \ ,
\label{endcommut}
\eeq
with $k$ and $T_{b'}$ as given after equation (\ref{conjid}). That is, in each model, the textured string varies only over some $SU(2)$ (or $SO(3)$) subgroup of $G$, in which $U(r,\varphi)$ assumes nontrivial values. Unless the generator $T_{h'}$ commutes with that entire subgroup --- thus being entirely insensitive to the string --- $T_{h'}$ will show Alice behavior, through charge violation at finite nonzero radius. For the
canonical Alice/nematic liquid crystal model of section
\ref{canonagain}, this means that charge associated with the single
unbroken generator $T_h = T_{h'} = T_z$ is violated at all finite
nonzero $r$.  For the $SO(6)$ model of section \ref{SON}, we have six
unbroken generators $T_{12}, T_{13}, J_1,K_1, T_{46}, T_{56}$ with
$T_h = J_1 = T_{23} + T_{45}$. All charges are violated at finite
nonzero radii, except the one associated with $K_1$. This charge must be
conserved: in our $SO(4)$ subgroup's $SU(2) \times SU(2)$ product structure,  the textured Alice string varies  over
the $J$-associated $SU(2)$ and commutes with the $K$-associated one.

The details of this charge violation are more subtle than the divergent case,
however; they consist not of a simple $Z_2$ charge flip $\ket{\pm} \
\rightarrow\ \ket{\mp}$. Instead an r-dependent rotation mixes, say, a
positively charged eigenstate into a linear combination of positive,
negative, and neutral eigenstates. For example, in our $SO(6)$ model of section
\ref{SON}, the unbroken generator $T_{h'} =
T_{12}$ has eigenstates which undergo
charge-violating Aharonov Bohm scattering at finite $r$. The 
$T_{12}$ charge eigenstates $ \ket{\pm}$ scatter to
\beq
\ket{\pm}' =
\sfrac{1}{2}\ (\ 1+ \cos\alpha  \ )\  \ket{\pm} 
+ \sfrac{1}{2}\ (\ 1 - \cos\alpha  \ )\  \ket{\mp} 
\ \pm\ \sfrac{i}{\sqrt{2}}\ \sin \alpha \ \ket{0}\ \ ,
\eeq
using the effective Wilson loop discussed above; where $\ket{0}$ is the
$r$-dependent neutral charge eigenstate 
\beq
\ket{0} = \cos (kF)\ \ket{e_3}\ +
\ \sin (kF)\ \ket{e_4}\ \ .
\eeq
Thus scattered charge
remains unchanged at the origin and spatial infinity; elsewhere, it
becomes a radially dependent admixture of variously charged
states. Note that, although this textured Alice string is a $Z_2$
string topologically, its charge violation is generally not $Z_2$: two
traversals around the string are not trivial, at most radii. Only at
the special ``midpoint'' radius, where $\alpha = \pi, kF = \pi/2$,
does a $Z_2$ scattering occur: $\ket{\pm} \rightarrow \ket{\mp}$. 

Similarly, for our canonical Alice/nematic liquid textured Alice
string of section \ref{canonagain}, $\ket{\pm}$ charge eigenstates of $T_h = T_z$ scatter to
\begin{eqnarray}
\ket{\pm}' &=&
\sfrac{1}{2}\ [\ 2\cos(2\alpha) + (\ 1-\cos(2\alpha)\ )\ \sin^2F \ ]\  \ket{\pm} \  -\ \sfrac{1}{2}\ (\ 1 - \cos(2\alpha)  \ )\ \ \sin^2F\ \ket{\mp} 
\nonumber\\ 
&& +\ \sfrac{1}{\sqrt{2}}\ \sin F \ 
\left[\ -\sin (2\alpha) \pm i \cos F\ (1-\cos(2\alpha))\ \right]\ \ket{0}\ \ .
\end{eqnarray}
Again we find nontrivial, but non-$Z_2$, mixing of the charge
eigenstate into states of different charge. At the string
``midpoint,'' $\alpha = \pi, F = \pi/2$, the charge is actually conserved, as
$\ket{\pm} \rightarrow \ket{\pm}$. Elsewhere, however, for $0<F<\pi$ (finite nonzero $r$), the charge eigenstates mix nontrivially.

We have thus shown that circumnavigation of the string at finite
radius violates charge conservation, for charges defined either
asymptotically or locally, in accordance with local unbroken
symmetries. Those charges associated with unbroken symmetries are
simply nonconserved, at finite radius. 

In the gauged or divergent global Alice string case, charge violation
occurs because local residual symmetries $h$ cannot be consistently
extended to singlevalued symmetries $h\,(\varphi)$. Here that is not
true --- the fact that the group map $U(r,\varphi)$ generating the
string is single-valued induces perfectly single-valued symmetries,
and symmetry generators $T_{h'}$, when parallel transported around the
string by (\ref{parallelt}). However, no set of associated conserved
charges can be defined, at finite radius. One might hope some
``hidden'' set of charge eigenstates, murkily connected with the local
unbroken symmetry, might emerge. However, the only candidates for such
conserved charge eigenstates are those simultaneous charge eigenstates
$\ket{q_h, q_{h'}, \ldots}$, for all $T_{h'}$ commuting with
$T_h$. These, at least, are conserved in circumnavigating the string,
acquiring only the r-dependent anyonic phase
$e^{-iq_h\alpha}$. However, any deviation from a circular path in
traversing the string violates these eigenstate's charge, for the same
reason that asymptotically defined charge becomes violated at finite
radius. That is, charge is nonconserved because radial transit causes
the charge eigenstates to change and become misaligned with the Wilson
loop $h (\,\alpha\,)$.

\subsubsection{For Majorana Fermions}

Understanding the Aharonov Bohm scattering for Majorana fermions in
the Benson-Manohar-Saadi model is slightly more complex. Here fermions
$\psi$ interact with the textured string
\beq \vev\ (r,\varphi)  \  = \ 
U\ (r,\varphi) \ \vev_o\  U^{T}\ (r,\varphi) \ \ 
 \  = \  h(\varphi/2)\ \ e^{-2iF(r)\,\sigma_{z(12) }}\ \ h^{-1}\ (\varphi/2)
\ \ 
\label{BMSvev}
\eeq
via the mass interaction $-m\bar{\psi}\ \phi\ \psi_c + {\it
h.c.}$ 
At first glance, this gives the mass eigenstates (\ref{BMSmasseig}),
with mass eigenvalues $m\, e^{\mp 2iF(r)}$, and with nonabelian Aharonov
Bohm phase cancelling the apparent angular variation of the mass
eigenstates. However, to get the masses and nonabelian Berry's phase
correct, we must look more closely at the bispinor $(\psi,\psi_c)$ and how it obtains its mass. We include a self-conjugate bare Majorana mass matrix $M$ for clarity. The full mass matrix is
\beq
\left(\ \begin{array}{cc} &M + m \vev\\ M + m \vev^\dagger& \end{array}\ \right)\ \ 
\eeq
acting on $(\psi,\psi_c)$. It has mass eigenstates 
\beq
\psi_{1\pm} = \frac{1}{\sqrt{2 \omega*\,\omega}}\ \left( \begin{array}{r} \omega\ \psi_+\ \chi \\
 \pm\ \omega^*\ \psi_+\ \chi\end{array}\ \right)
\ ,
\ 
\psi_{2\pm} =  \frac{1}{\sqrt{2 \omega*\,\omega}}\ \left( \begin{array}{r} \omega^*\ \psi_-\ \chi \\
 \pm\ \omega \ \psi_-\ \chi\end{array}\ \right)
\label{BMSfullmasseig}
\eeq
where
\beq
\omega = \left(\ M + m\, e^{-2iF} \ \right)^{1/2}\ \ ,
\eeq
$\psi_\pm$ are the mass eigenstates (\ref{BMSmasseig}) in flavor space and $\chi$ is the self-conjugate field $\psi + \psi_c$ in spin space. These have associated, spatially varying mass eigenvalues
\beq
\lambda_\pm = \pm\ \omega^*\,\omega = \pm\ \sqrt{\ M^2 + m^2 + 2Mm\cos(2F)
\ }\ \ ,
\eeq
with degeneracies between $\psi_{1+}$ and  $\psi_{2+}$, $\psi_{1-}$ and  
$\psi_{2-}$.

We now calculate the nonabelian Aharonov Bohm phase
(\ref{nonaBphaseagain}) due to these degeneracies. First note that
$\partial_\varphi$ acting on any of the spatially varying mass
eigenstates (\ref{BMSfullmasseig}) is just $(-i\sigma_{y(12) }/2, -i\sigma_{y(12) }/2)$ acting on that
eigenstate, as the $\varphi$-variation of each eigenstate is simply
the $h(\varphi/2)$ prefactor inherited from flavor mass eigenstates
(\ref{BMSmasseig}). And $i\sigma_{y(12) }$ acts on the flavor mass eigenstates
(\ref{BMSmasseig}) by sending $\psi_- \ \rightarrow\ \psi_+,\ \psi_+ \
\rightarrow\ -\, \psi_-$. This affects the full mass eigenstates
(\ref{BMSfullmasseig}) as follows:
\beq
(i\sigma_{y(12) }, i\sigma_{y(12) })\ :\ \psi_{a\pm} \ \rightarrow\ - \epsilon_{ab3}\ \psi_{b\pm}^*\ \ .
\eeq
Thus our  nonabelian Aharonov Bohm gauge field $A_\varphi$ mixes only degenerate mass eigenstates, with matrix elements
\begin{eqnarray}
(A_\varphi)_{ab,\,\pm} &=& 
\langle\ \psi_{b\pm}\ (r, \varphi)\ | \ \partial_\varphi \ |\ \psi_{a\pm}\ 
(r, \varphi)\  \rangle
\quad\quad =\quad\quad 
\frac{1}{2}\ \epsilon_{ab3}\ \langle\ \psi_{b\pm}\  |\ \psi_{b\pm}^*\ 
 \rangle\nonumber\\[4pt]
&=&\frac{1}{4 \omega^*\,\omega }\ \epsilon_{ab3}\ \left(\ \omega^2 + {\omega^*}^2\ \right) \nonumber\\[4pt]
&=&\epsilon_{ab3}\ \frac{M + m\cos(2F)}{2\sqrt{\ M^2 + m^2 + 2Mm\cos(2F)
\ }}\ \ .
\end{eqnarray}
This gives for the nonabelian Wilson line
\beq
W(r,\varphi) = h^{-1}\ \left(\  \frac{M + m\cos(2F)}{2\sqrt{\ M^2 + m^2 + 2Mm\cos(2F)
\ }}\ \varphi\ \right)\ \ ,
\eeq
where $h(\varphi)$ is now understood as the direct product $(e^{-i\varphi\sigma_{y(12) }}, \ e^{-i\varphi\sigma_{y(12) }})$ acting on the bispinor $(\psi, \psi_c)$. Thus the $\varphi = 0$ mass eigenstates (\ref{BMSfullmasseig}) are acted on by
\beq
h(\pi)\ W(r,2\pi) = h\, \left(\ \left(\ 1 - \frac{M + m\cos(2F)}{\sqrt{\ M^2 + m^2 + 2Mm\cos(2F)
\ }}\ \right)\ \pi\ \right) 
\eeq
in traversing the string. This gives single-valued mass eigenstates at
spatial infinity and the origin, where $F$ assumes the values 0 and
$\pi/2$ respectively. Elsewhere, mass eigenstates become multivalued in
traversing the string. In the case where the bare Majorana mass $M$ is
zero, our effective Wilson loop is
\beq
h(\pi)\ W(r,2\pi) = h\, \left(\ \left(\ 1 - \cos(2F)\ \right)\ \pi\ \right)\ \ . 
\eeq

We note that this is identical to the effective $r$-dependent Wilson
loop $h(\,\alpha\,)$ we obtained for the Dirac case above. And
identical results ensue. Note that, as in the Dirac case, local
symmetry generators vary radially:
\beq
T_{h',r} = e^{-iF(r)\,\sigma_{z(12) }}\ T_{h'}\ e^{+iF(r)\,\sigma_{z(12) }}\ \ ,
\eeq
due to the symmetry transformation rule $\vev \ \rightarrow\ g\ \vev\
g^T$. The mismatch --- that is, noncommutativity --- between the
Wilson loop $h(\,\alpha\,)$, generated by $T_h$, and the radially
twisted local unbroken generators $T_{h',r}$, is why a local observer
at finite $r$ perceives charge violation, when charge associated with unbroken
symmetries at $r$ circumnavigate the string. Similarly,
the full mass eigenstates (\ref{BMSfullmasseig}) vary radially. For
$M=0$, in fact, they vary as if acted on by the flavor space operator
$(\ e^{-iF(r)\,\sigma_{z(12) }},\ e^{+iF(r)\,\sigma_{z(12) }}\
)$. This causes the same noncommutativity problem as in the Dirac
case, when asymptotic charges approach finite radius, tracking
radially varying mass eigenstates; then circumnavigate the string,
acquiring the mismatched Wilson loop $h(\,\alpha\,)$; then return to
spatial infinity. As in the Dirac case, for either thought experiment
--- circumnavigation of the string at finite radius by charges defined
either locally, or brought in from infinity --- charges associated
with the asymptotic generator $T_{h'}$ are violated whenever
\beq
\commut{\ T_{h'}, \ \left(\ \cos\ [2F(r)]\ \sigma_{y(12) }  \ \ + \ \ \sin\ [kF(r)]\ \sigma_{x(12) }\ \right)\ }  \ \ket{q_{h'}}\ \ne 0\ \ ,
\eeq
following analysis similar to that preceding equation (\ref{endcommut}) for the Dirac case. This means in particular, for the $N=3$ BMS string of section \ref{Maj}, that all charges are violated at finite radius.

\section{Skyrmions as Twisted Textured Alice Strings }\label{twist}

We now show that the textured Alice strings constructed here can be
twisted to form topologically nontrivial skyrmions --- that is, point
defects with nontrivial $\pi_3(G/H)$. The twist we consider is the following: bend the string into a loop, with angle $\theta$ along the string core $2\pi$-periodic. As $\theta$ increases along the loop, twist the textured string by starting its $h$-rotation at increasing offset angle $\varphi_o = \theta$:
\beq
U\ (r,\ \theta + \varphi) =  h(\ (\varphi + \theta)/2\ )\ \ e^{-iF(r)\, T_b}\ \ h(\ (\varphi + \theta)/2\ )\hspace{0.25in}
\mbox{where} \quad F(r) \rightarrow \left\{ \begin{array}{ll}0&r\rightarrow\infty\\F_o&r\rightarrow 0
\end{array}\right. \ \ ,
\eeq
giving strings
\begin{eqnarray}
 \vev\ (r,\ \theta + \varphi)  &= &h(\ (\varphi+\theta)/2\ )\ \ e^{-iF(r)\, T_b}\ \ \vev_o\ \ e^{+iF(r)\, T_b}\ \ h^{-1}\ (\ (\varphi+\theta)/2\ )\hspace{0.25in}\mbox{Dirac case}\nonumber\\
\vev\ (r,\ \theta + \varphi)  &=&
  h(\ (\varphi+\theta)/2\ )\ \ e^{-2iF(r)\,\sigma_{z(12) }}\ \ h^{-1}\ (\ (\varphi+\theta)/2\ )
\hspace{0.85in}\mbox{Majorana case}\ \ ,
\label{twiststrings}
\end{eqnarray}
generalizing the untwisted strings of Equations (\ref{genUagain}),
(\ref{form2again}), and (\ref{BMSvev}). Note that this twist is
distinct from the gauged monopole case: for the gauged monopole case,
the {\em plane} of the string's internal space rotation rotates as we
traverse the loop, locked to the loop angle $\theta$ to form a
hedgehog configuration. Here, the textured string always rotates in
the same internal space plane, assuming the same vacuum values; the
entire configuration just shifts in $\varphi$ by an offset angle
locked to $\theta$.

This offset of identical string configurations is essential in
establishing nontrivial homotopy $\pi_3$. For each of our models
varies nontrivially over, essentially, an $SU(2)\ \rightarrow\ SO(2)$
subspace of the model's symmetry-breaking pattern $G \rightarrow H$.
For our $SO(6)$ model of section \ref{SON}, our model varies over an
$SO(4)\ \sim\ SU(2)\ \times\ SU(2)$ subspace, where $SU(2)\ \times\
SU(2) \ \rightarrow U(1) \ \times\ U(1)$; note that our model varies
only over one $SU(2) \ \rightarrow\ U(1)$ factor, associated with the
$J$-subalgebra. The canonical Alice model of section \ref{canonagain}
is explicitly a symmetry breaking $SO(3) \ \rightarrow\
O(2)$. Finally, the BMS model of section \ref{Maj} varies over an
$SU(2)\ \rightarrow\ SO(2)$ subspace of the symmetry-breaking $SU(N)\
\rightarrow\ SO(N)$.

In all these cases, then, we can exploit the isomorphism $SU(2)/SO(2)\
\sim\ S^2$ to identify the twisted loop's $\pi_3$ index with its Hopf
number, the linking number of any two fibers of constant $\vev$ in
physical space. \cite{wuzee} Our twisted textured strings
(\ref{twiststrings}) were constructed to be exactly $2\pi$-periodic in
the angle $\varphi+\theta$ at finite $r$, and angle-independent at
$r=0$. Thus we obtain one fiber by taking $r=0$ along the core of the
loop, and a second by taking finite $r=r_o$, with $\theta + \varphi =
0$. This second fiber, or constant value for $\vev$, occurs at angle
$\varphi = -\theta$ as we traverse the string loop taking $\theta$
from 0 to $2\pi$. The two fibers are drawn below in Figure 2. Note
that endpoints in $\theta$ and $\varphi$ are identified. The linking
number is exactly one, showing that the twisted textured Alice string
is a fundamental skyrmion. Thus, as in the gauged case, twisted
textured Alice strings form fundamental point defects. Unfortunately,
here the argument is not generic, as in the gauge case; it relies
specifically on the Hopf map for the vacuum submanifold $SU(2)/U(1)$.

\vspace{12pt}

\epsfig{file=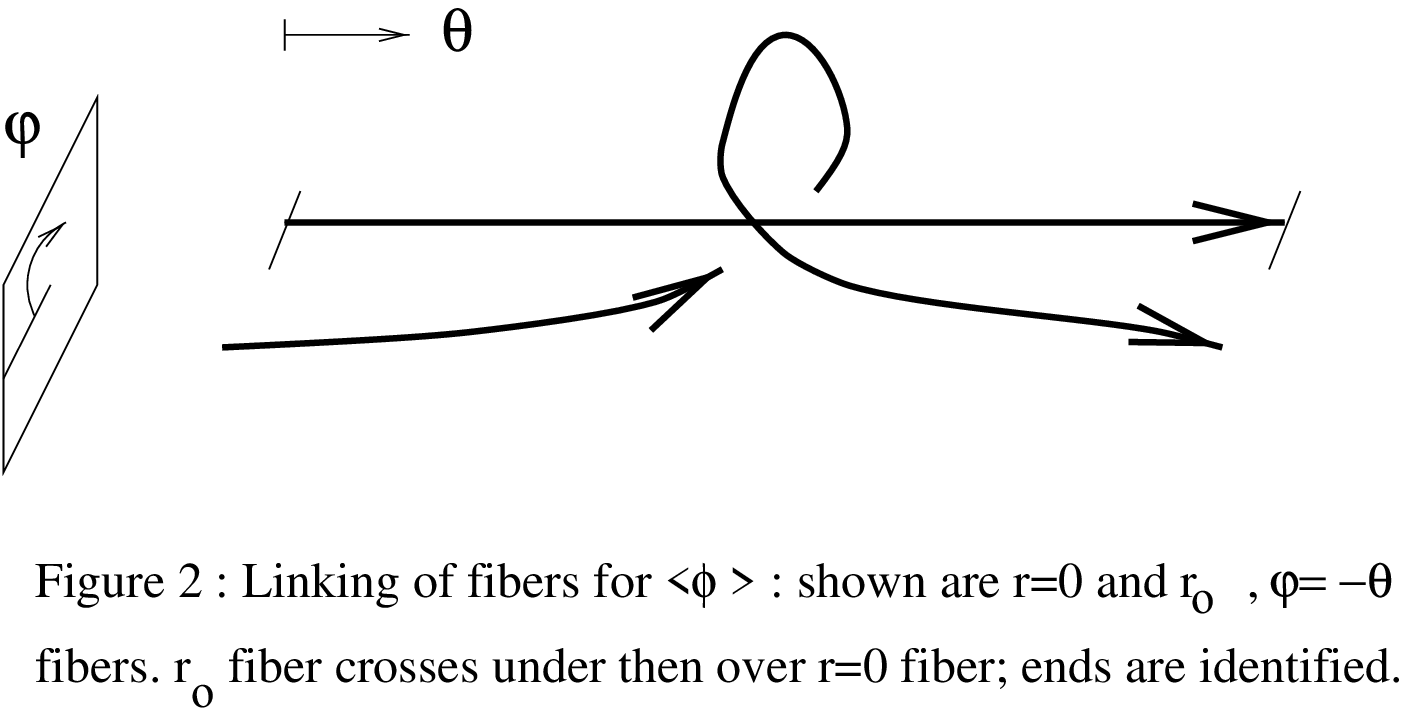, width=5.5in}

\section{Conclusions}\label{conclude}

We have demonstrated Alice behavior for global strings of both types:
divergent global strings with nontrivial $\pi_1 (G/H)$, and finite
tension textured strings with nontrivial $\pi_2(G/H)$. In both cases,
point defects emerge as twisted strings.  Charge violation also
emerges in both cases: in the first case, the global string inherits
its Alice characteristics from its gauged counterpart; while in the
second, charge violation is generic, at finite radius. Specifically,
for the divergent global string, charge violation occurs for all
generators that fail to commute with $W(2\pi)\ U(2\pi)$, where
$U(2\pi)$ is the Wilson loop of the associated gauged string, and
$W(2\pi)$ is a rare correction due to the nonabelian Aharonov Bohm
effect. In the textured case, charge violation occurs for all
generators which interact with the string at all, as a direct
consequence of the nonabelian Aharonov Bohm effect. These textured
strings are associated with topologically nontrivial loops in $H$,
$h(\varphi)$. We showed that the nonabelian Aharonov Bohm effect
determines an effective Wilson loop for the string at finite radius of
$h(\alpha)$, where $\alpha$ ranges from 0 at infinite radius to $2\pi$
at $r = 0$. At finite radius, a mismatch between this Wilson loop, and
locally defined unbroken symmetry generators, leads to generic and
observable charge violation.

\acknowledgements This work was supported by NSF grant PHY-9631182,
and by the University Research Committee of Emory University. I thank
Tom Imbo for useful discussions, particularly on subtleties of the
Hopf map.

\end{document}